\documentstyle{article}
\input math_macros
\font\tenbf=cmbx10
\font\tenrm=cmr10
\font\tenit=cmti10
\font\elevenbf=cmbx10 scaled\magstep 1
\font\elevenrm=cmr10 scaled\magstep 1
\font\elevenit=cmti10 scaled\magstep 1

\renewenvironment{thebibliography}[1]
 { \elevenrm
   \begin{list}{\arabic{enumi}.}
    {\usecounter{enumi} \setlength{\parsep}{0pt}
     \setlength{\itemsep}{3pt} \settowidth{\labelwidth}{#1.}
     \sloppy
    }}{\end{list}}

\begin{document}
\renewcommand{\thefootnote}{\alph{footnote}}
\begin{center}{{\tenbf
               Parton-Parton Elastic Scattering and Rapidity Gaps\\
               \vglue 3pt
               at SSC and LHC Energies   \\}
\vglue 1.0cm
{\tenrm Vittorio Del Duca \\}
\baselineskip=13pt
{\tenit Stanford Linear Accelerator Center\\}
\baselineskip=12pt
{\tenit Stanford University, Stanford, California\\}
\vglue 0.8cm
{\tenrm ABSTRACT}}
\end{center}
\vglue 0.3cm
{\rightskip=3pc
 \leftskip=3pc
 \tenrm\baselineskip=12pt
 \noindent
The theory of the perturbative pomeron,
due to Lipatov and collaborators, is used to compute the
probability of observing parton-parton elastic scattering and rapidity
gaps between jets in hadron collisions at SSC and LHC energies.
\vglue 0.6cm}
{\elevenbf\noindent 1. Introduction}
\vglue 0.4cm
\baselineskip=14pt
\elevenrm

At the SSC and LHC hadron colliders events predicted by the Standard Model,
like Higgs-boson production
via electroweak boson fusion, will be experimentally accessible.
A characteristic signature of this process is that in the
$t$ channel no color is exchanged between the scattering hadrons,
the color exchange being confined to the fragmentation region between
the struck and spectator partons$^1$. On a Lego plot in
azimuthal angle and rapidity, the signal will present, at the parton level,
a rapidity gap between the
struck partons$^2$. In order, though, for the gap to
survive at the hadron level, it is necessary that no color radiation is
exchanged between the spectator partons in the hadronization process.
Bjorken$^2$ has first defined and estimated the survival of the
rapidity gap in the presence of soft spectator interactions.
A study of the event characteristics of Higgs-boson production via
electroweak-boson fusion has then been recently undertaken$^3$.

The Higgs boson may be produced also via gluon-gluon fusion. This
will usually have color all over the $t$ channel, since the gluons, on their
way to fuse together, will emit gluon radiation. A fraction of events
in this process, though, may simulate Higgs-boson production via
electroweak-boson fusion even in its
characteristic signature, namely the Higgs boson may be produced by the
fusion of two pairs of gluons in color singlet configurations$^{2,4}$.
Then no color is exchanged in the $t$ channel between the struck
partons. To understand the dynamics of these background events, it is better to
undertake the
propedeutic study of hadron-hadron scattering with exchange in the $t$
channel of a pair of gluons in a color singlet configuration. Such a
study can be already done experimentally at the energies of the Tevatron
collider$^5$, and indeed the first data on rapidity gaps in hadron
collisions starts being available$^6$.
\vglue 0.6cm
{\elevenbf\noindent 2. Rapidity Gaps}
\vglue 0.4cm
In this talk I illustrate a way of computing the probability of
observing parton-parton elastic scattering and rapidity gaps between
jets in hadron collisions at very high energies$^7$, and use it to make
predictions on rapidity-gap production at SSC and LHC energies.
In     order    to    obtain    quantitative
predictions   of jet production in the very high energy limit
and separate it from the  uncertainty
involving  the small  $x$  dependence  of parton
distributions$^8$, Mueller and
Navelet$^9$ proposed to measure the two-jet
inclusive cross section in hadron
collisions by tagging two jets at a large rapidity interval $y$ and with
transverse  momentum  of order $m$.
These tagging jets are produced in a nearly forward scattering of gluons or
quarks with  large center-of-mass energy $\sqrt{\hat s}$.
Lipatov and collaborators$^{10-13}$ (BFKL) have shown that,
in this regime, the rapidity interval $y=\ln({\hat s}/m^{2})$
between the scattered partons is filled in by the radiation of additional
gluons, roughly uniformly spaced in rapidity, all with transverse
momenta of order $m$.
The BFKL theory systematically corrects the lowest-order QCD result
by summing the leading logarithms of $\hat s$. The result is to
replace the gluons exchanged in the $t$ channel with effective,
reggeized gluons, with an infrared-sensitive trajectory$^{11}$.
Then one uses this resummed, effective
gluon exchange to compute the elastic amplitude in the Regge limit
$\hat s \gg -\hat t$ with color singlet exchange in the $t$-channel.
This is known as the BFKL pomeron$^{11-13}$.
The imaginary part of the forward amplitude is the
parton-parton total cross section.
To leading order in rapidity, the parton-parton total
cross section and the related
2-jet inclusive cross section exhibit the energy
dependence $\exp[(\alpha_{P}-1)y]$ with
\begin{equation}
\alpha_{P}=1+4\ln2{\alpha_s C_A \over \pi},
\end{equation}
where $C_A = N_c = 3$ is the number of colors in QCD.

The elastic scattering amplitude with color singlet exchange in the
$t$ channel is a higher order
($\alpha_s^4$) process$^{13-14}$ but with energy dependence
$\exp[2(\alpha_{P}-1)y]$, and it leads to a final state which, at the
parton level, contains two jets with a rapidity gap in gluon production
between them. Some fraction of these states may produce the
experimental signatures of a large rapidity gap in secondary
particle production.

To understand the relation between rapidity gaps
in hard-gluon and hadron production, we must discuss the potential
backgrounds to these signals at the parton and hadron level.
To analize the parton-level background, assume that we cannot
detect partons with transverse momentum smaller
than a fixed parameter $\mu$. In this case, there is an additional
contribution to elastic scattering from color octet exchange in the $t$
channel. According to BFKL, this proceeds via the exchange of a
reggeized gluon, which contains all the leading virtual
radiative corrections and it has the form of a Sudakov form factor.
The parameter $\mu$ fixes the scale below which soft gluon radiation is
allowed. As $\mu \rightarrow 0$, the contribution of the color octet
exchange vanishes, since it is impossible to have scattering with
exchange of a gluon, without allowing for the emission of soft gluon
radiation.

In order to use perturbative QCD, the parameter $\mu$ must be larger
than $\Lambda_{QCD}$. Thus we have two options:
first, we can consider $m \gg \mu \gg \Lambda_{QCD}$, that is, we define a
rapidity gap to be present if there are no jets of transverse momentum
larger than $\mu$ between the tagging
jets. We will call this case $quasi\,\, elastic\,\,
scattering$, since it allows gluon radiation below the scale $\mu$.
The ratio $R$ of the quasi-elastic to the total
cross section is given by
\begin{equation}
R(\mu) = {\sigma_{singlet} + \sigma_{octet} \over \sigma_{tot}}.
\end{equation}
where all the cross sections in (2) have been convoluted with the
appropriate parton distributions.
Alternatively, we can consider
$\mu = O(\Lambda_{QCD})$. Then at the parton level the color octet
exchange is strongly suppressed, and only the color singlet exchange
contributes to the cross section for producing rapidity gaps.

At the hadron level, the interaction between spectator
partons may produce hadrons across the rapidity interval,
spoiling the rapidity gap. Thus in
order to compute the cross section for producing a rapidity gap at the
hadron level, we need a non-perturbative model which describes the hadron
interaction and estimates the survival of the rapidity gap in the presence
of soft spectator interactions$^2$.
The rapidity-gap survival probability $<S^2>$ is defined
as the probability that in a scattering event no other interaction
occurs beside the hard collision of interest.
$<S^2>$ is expected to depend on the hadron-hadron center of mass
energy, but only weakly on the size of the rapidity gap.
Then to obtain the probability of a scattering
event with a large rapidity gap at the hadron level, we must compute
the ratio $R$ at $\mu = 0$, that is, using only the singlet elastic
cross section, and multiply it by the survival probability $<S^2>$:
\begin{equation}
R_{gap} = <S^2> R(\mu = 0).
\end{equation}
In this contribution, we compute $R(\mu)$ at the
SSC and LHC center-of-mass energies $\sqrt{s}$ of 40 TeV and 16 TeV
respectively, and at different
values of the minimum transverse momentum of the tagging jets $m$ and
the elastic scale $\mu$.
\vglue 0.6cm
{\elevenbf\noindent 3. Jet Cross Sections}
\vglue 0.4cm
We consider the scattering of two hadrons of
momenta $k_A$ and $k_B$ in the center-of-mass frame
and we imagine to  tag two jets at the extremes
of the Lego plot, with the rapidity interval between them filled
with jets. The tagging jets
can be characterized by their transverse momenta $p_{A\perp}$ and
$p_{B\perp}$ and by their rapidities $y_A$ and $y_B$. The inclusive
cross section for producing two tagging jets with transverse momenta
greater than a minimum value $m$ is then$^9$
\begin{eqnarray}
& &{d\sigma \over dy d\bar{y}}(AB\rightarrow j(x_A) j(x_B) + X) = \nonumber\\
& & \int dp_{A\perp}^2 dp_{B\perp}^2 \prod_{i=A,B}
   \biggl[G(x_i,m^2) + 4/9 \sum_f [Q_f(x_i,m^2) +
   \bar Q_f(x_i,m^2)]\biggr] {d\hat\sigma_{tot} \over dp_{A\perp}^2
   dp_{B\perp}^2}
\end{eqnarray}
where $x_A \simeq e^{y_A} p_{A\perp}/\sqrt{s}$,
$x_B\simeq e^{-y_B} p_{B\perp}/\sqrt{s}$ are the light-cone momentum fractions
of the tagging
jets with respect to their parent hadrons, $y = |y_A-y_B|$ is the
rapidity difference and $\bar{y} = (y_A+y_B)/2$ is the rapidity boost,
$\hat s = 2\, k_A \cdot k_B x_A x_B$ is the parton-parton squared
center-of-mass energy, and
\begin{equation}
{d\hat\sigma_{tot} \over dp_{A\perp}^2 dp_{B\perp}^2} =
{(\alpha_s C_A)^2 \over 2 p_{A\perp}^3 p_{B\perp}^3}
\int_0^{\infty} d\nu e^{\omega(\nu)y} \cos\left( \nu\ln {p_{A\perp}^2 \over
p_{B\perp}^2} \right)
\end{equation}
is the BFKL total cross section for gluon-gluon
scattering within an impact distance of size $1/m$, and
\begin{equation}
\omega(\nu) = {2 \alpha_s C_A \over\pi}\left[ \psi(1) - \Re\psi
   ({1\over 2} +i\nu) \right],
\end{equation}
with $\psi(z)$ the logarithmic derivative of the Gamma function.
In eq. (4) we use
the large-$y$ effective parton distribution functions$^{15}$, computed
at the factorization scale $Q^2 = m^2$.

The high-energy elastic cross section for two tagging jets, with color
singlet exchange in the $t$ channel, is
\begin{equation}
{d\sigma_{sing} \over dy d\bar{y}}(AB\rightarrow j(x_A) j(x_B)) =
\int d\hat{t} \prod_{i=A,B}
\biggl[G(x_i,m^2) + (4/9)^2 \sum_f [Q_f(x_i,m^2) + \bar Q_f(x_i,m^2)]\biggr]
{d\hat\sigma_{sing} \over d\hat{t}},
\end{equation}
where $\hat t \simeq -p_{\perp}^2$, with $p_{\perp}$ the transverse
momentum of the tagging jets. The gluon-gluon elastic scattering
cross section, with the tagging jets collimated and with minimum
transverse momentum $m$, is$^{14}$
\begin{equation}
{d\hat\sigma_{sing} \over d\hat{t}} = {(\alpha_s C_A)^4 \over 4\pi
\hat{t}^2} \left(\int_{-\infty}^{\infty} d\nu {{\nu}^2\over \left(
{\nu}^2+{1\over
4} \right)^2} e^{\omega(\nu)y} \right)^2.
\end{equation}
Since two reggeized gluons are involved in the color singlet
exchange in the $t$ channel, in keeping into account in (7) the possibility
that the scattering is initiated by quarks we obtain the suppression
factor $(4/9)^2$. The background to the color singlet exchange comes
from the exchange of a reggeized gluon. This contribution is given by
\begin{equation}
{d\sigma_{octet} \over dy d\bar{y}}(AB\rightarrow j(x_A) j(x_B) ) =
 \prod_{i=A,B}
   \biggl[G(x_i,m^2) + 4/9 \sum_f [Q_f(x_i,m^2) +
   \bar Q_f(x_i,m^2)]\biggr] {d\hat\sigma_{oct} \over d\hat{t}},
\end{equation}
where the gluon-gluon
elastic scattering cross section in the color octet channel is
\begin{equation}
{d\hat\sigma_{oct} \over d\hat{t}} = {\pi (\alpha_s C_A)^2 \over 2\hat
t^2} \exp\left( -{\alpha_s C_A \over \pi} \, y \, {1\over
\sqrt{1+4\mu^2/p_{\perp}^2}} \ln{\sqrt{1+4\mu^2/p_{\perp}^2} + 1 \over
\sqrt{1+4\mu^2/p_{\perp}^2} - 1} \right).
\end{equation}
For $m \gg \mu$ the exponential reduces to$^{14}$ $\exp(- \alpha_s C_A / \pi
\ln(p_{\perp}^2/\mu^2) \, y)$ and has the typical form of a Sudakov form
factor.
As $\mu \rightarrow 0$, or $y$ becomes large, the contribution of the
octet to the gluon-gluon elastic cross section vanishes.
\vglue 0.6cm
{\elevenbf\noindent 4. The Numerical Evaluation of the Ratio $R(\mu)$}
\vglue 0.4cm
$R(\mu)$ is the probability of having elastic scattering at the parton
level, as defined in (2), and is obtained by summing (7) and (9),
and dividing by (4). To evaluate it, we scale the running coupling
constant from $\alpha_{s}(m(Z))=0.12$ using the 1-loop evolution with 5
flavors, and use the CTEQ set-5 parton distribution functions$^{16}$.
We plot $R(\mu)$ at LHC and SSC energies as a function of $y$, at rapidity
boost $\bar y$ = 0, and with
$m$= 30, 60, 100 GeV and elastic scale $\mu$ = 0, 0.5, 2, 5, 10 GeV.
The rapid growth of $R(\mu)$ at the kinematical upper boundary in $y$ is
due to the energy dependence of the pomeron trajectory (1)
enhanced by the scaling behavior at $x$ near 1 of the
distribution functions integrated over transverse momentum.
\vglue 19.0cm
\begin{center}{\baselineskip=12pt The ratio $R(\mu)$ as a function of
$y$, at $\bar y$ = 0, \\ at $\sqrt{s}$= 16 TeV in the left column and
$\sqrt{s}$= 40 TeV in the right column.}\end{center}
The growth of $R(\mu)$ due to the pomeron trajectory$^7$ only is more
apparent in the plots with $m$ = 30 GeV where the largest kinematically
accessible values of $y$ can be probed.
At $\mu = 0$, the octet exchange does not contribute to $R(\mu)$.
The value of $R(\mu = 0)$, multiplied by the survival probability $<S^2>$,
gives the probability of having a collision with a rapidity gap in
secondary particle production (3). Since $<S^2>$ is estimated in ref. 2
to be $\simeq$ 0.1 and in ref. 17 to be in the range of 0.05 to 0.2,
we expect that, at the LHC and SSC, a few tenths percent of events with
tagging jets will show rapidity gaps. $<S^2>$, though, is expected to
decrease with $\sqrt{s}$, so fewer rapidity-gap events should be found at
LHC and SSC energies than at Tevatron energies, at comparable values of
$y^{18}$.
The probability of finding a gap
increases with the rapidity interval between the tagging jets. This
prediction is peculiar of the radiative corrections to $R(\mu)$,
since $R(\mu)$ at the lowest order in $\alpha_S$ does not depend on
$y$.

Since all of the analysis above is in the leading logarithmic approximation,
there is ambiguity in the choice of the proper scale in rapidity for
which this analysis is valid, and so the exact value of the
normalization and thus of $R(\mu)$ cannot be determined precisely.
However, the slope of the curves in the
asymptotic regime is free from this scale uncertainty and thus the
experimental measurement of the ratio $R(\mu)/R_{gap}$ in the large
rapidity-gap regime should give
us an unambiguous determination of the survival probability $<S^2>$.
\vglue 0.5cm
Work supported by the Department of Energy, contract DE-AC03-76SF00515.
I wish to thank Bj Bjorken, Jerry Blazey, Andrew Brandt,
Jeff Forden, Michael Peskin, Carl Schmidt, Michael Sotiropulos,
George Sterman, Wai-Keung Tang and Harry Weerts for many stimulating
discussions
and
suggestions.

\end{document}